\documentclass[12pt]{article}
\usepackage{amssymb,amsmath,esint,titlesec,amscd}
\titleformat*{\section}{\large\bf}
\titleformat*{\subsection}{\normalsize\bf}

\begin{document}

%%%%%%%%%%%%%%%%%%%%%%%%%%%%%%%%%%%%%%%%%%%%%%%%%%%%%%%%%%%%%%%%%%%%%%%%%%%%%%%

\begin{titlepage}

\setlength{\baselineskip}{24pt}

                               \vspace*{0mm}

                             \begin{center}
                             
{\Large\bf  The $\mathbf{\tau}_\mathbf{q}$-Fourier transform: covariance and uniqueness}

                            \vspace*{3.5cm}

              \large\sf  NIKOLAOS  \  \  KALOGEROPOULOS $^\dagger$\\

                            \vspace{0.2cm}
                            
 \normalsize\sf Center for Research and Applications \\
                                  of Nonlinear Systems \ (CRANS),    \\
   University of Patras, \  Patras 26500, \ Greece.                        \\

                            \end{center}

                            \vspace{3.5cm}

                     \centerline{\normalsize\bf Abstract}
                     
                           \vspace{3mm}
                     
\normalsize\rm\setlength{\baselineskip}{18pt} 

We propose an alternative definition for a Tsallis entropy composition-inspired Fourier transform, 
which we call ``$\tau_q$-Fourier transform". 
We comment about the underlying  ``covariance" on the set of algebraic fields that motivates its introduction. 
We see that the definition of the $\tau_q$-Fourier transform is automatically invertible in the proper context. 
Based on recent  results in Fourier analysis, it turns that the $\tau_q$-Fourier transform is essentially unique 
under the assumption of the exchange of the point-wise product of functions with their convolution.
  
                           \vfill

                                                                         
                             \vfill

\noindent\rule{9cm}{0.2mm}\\  
   \noindent $^\dagger$ \small\rm Electronic address: \ \  \ {\normalsize\sf   nikos.physikos@gmail.com}\\

\end{titlepage}
 
%%%%%%%%%%%%%%%%%%%%%%%%%%%%%%%%%%%%%%%%%%%%%%%%%%%%%%%%%%%%%%%%%%%%%%%%%%%%%%%%%%%

\setlength{\baselineskip}{18pt}

\section{Introduction}

In this  note, we introduce the \ $\tau_q$-Fourier transform (whose name may stand for something along the lines of: 
``Tsallis entropy composition properties-induced Fourier transform of entropic index $q$"), as an alternative, but not equivalent, 
definition to  the $q$-Fourier transform.  Our proposal employs different  generalized sums and products than those on which 
the $q$-Fourier transform definition and properties rely. To make the exposition somewhat self-contained, 
we recall a few pertinent facts from previous works. \\

The Tsallis entropy \ $\mathcal{S}_q$ \ for a discrete set of 
outcomes/observations labelled by \ $j\in I$ \  with corresponding probabilities \ $p_j$ \ was defined by \cite{Ts,T-book}  
\begin{equation}
      \mathcal{S}_q [\{ p_j \}] \ = \ -k_B \cdot \frac{1}{1-q} \left\{ 1 - \sum_{j\in I} p_j^q  \right\}
\end{equation}
where \ $q\in \mathbb{R}$ \ is the entropic parameter, and the Boltzmann constant \ $k_B$ \ will be set equal to one from now on.  
Then, for conventionally defined independent outcomes \ $A, B$, \ namely ones satisfying \
$p[A\cup B] = p_A \cdot p_B$, \  one immediately sees that  the Tsallis entropy \ $\mathcal{S}_q$ \  
of the combined system \ $A\cup B$ \  is given by
\begin{equation}
     \mathcal{S}_q [p_{A\cup B}] \ = \ \mathcal{S}[p_A] + \mathcal{S}[p_B] + (1-q) \mathcal{S}[p_A] \mathcal{S}[p_B] 
\end{equation}
This suggests the introduction of the generalized sum \cite{T-book}
\begin{equation}
   x \oplus_q y = x+y + (1-q)xy
\end{equation}
in order to make (2) look more symmetric/covariant
\begin{equation}
      \mathcal{S}[p_{A\cup B}] \ = \ \mathcal{S}[p_A] \oplus_q \mathcal{S}[p_B]  
\end{equation}
Following (1), one subsequently defines the $q$-logarithm as \cite{T-book}
\begin{equation} 
      \log_q x \ = \ \frac{x^{1-q} -1}{1-q}, \hspace{5mm} x>0
\end{equation}
Its inverse is the $q$-exponential function \ $e_q(x)$, \ which turns out to be \cite{T-book}
\begin{equation}
         e_q(x) \ = \ [1+(1-q)x]_+ ^\frac{1}{1-q}
\end{equation}
where \  $(x)_+  = \max\{ x, 0 \}$. \ By introducing the  \ $q$-product \cite{NLeMW, Borges} 
as a variation of the well-known \ $l^p$ \ sum of Analysis, namely by 
\begin{equation} 
     x\otimes_q y \ = \ (x^{1-q} + y^{1-q} - 1)^\frac{1}{1-q} _+
\end{equation}
one recovers the familiar ``composition"  properties for the $q$-logarithm   
\begin{equation}
    \log_q (xy) \ = \  \log_q(x) \oplus_q \log_q(y), \hspace{15mm} \log_q (x\otimes_q y) \ = \ \log_qx + \log_q y
\end{equation}
and for the $q$-exponential 
\begin{equation}
   e_q(x\oplus_q y) \ = \ e_q(x) e_q(y), \hspace{20mm} e_q(x+y) \ = \ e_q(x) \otimes_q e_q(y) 
\end{equation}
The  definitions (5), (6) can be extended by interpreting them as the principal values of the corresponding 
complex functions  for the case \ $x, y \in \mathbb{C}$. \  Relying on the two generalized operations (3), (7), 
\cite{UT1, UTS, UTGS} define the $q$-Fourier transform \ $[\mathcal{F}_q(f)](\xi )$, \ for \ $q\in (1,3)$, \ for a function 
$f:\mathbb{R} \rightarrow \mathbb{R}_+$ with support $\Omega \subset \mathbb{R}$,  by the Lebesgue integral
\begin{equation}
  [\mathcal{F}_q(f)](\xi ) \ = \  \int_{\Omega} e_q (ix\xi ) \otimes_q f(x) \  dx  
\end{equation}  
where \ $i= \sqrt {-1}$. \ For a discrete set of functions \ $f_l: \mathbb{R} \rightarrow \mathbb{R}_+$ \ where \ $l \in I$, the definition is 
 \begin{equation}
    [\mathcal{F}_q(f)](\xi ) \ = \ \sum_{l\in I} e_q(i l \xi ) \otimes_q f_l
 \end{equation}
Since its introduction, several issues have arisen regarding this $q$-Fourier transform, 
\cite{Hil1, Hil2, UT2, PR1, JT, JTC, PR2, PR3}. 
Among them, one could single out its invertibility (10), (11) which has not been completely satisfactorily resolved, in our opinion. The
$q$-Fourier transform was introduced in order to be able to generalize the classical Central Limit Theorem valid for independent 
identically distributed random variables, to correlated random variables but in a particular way which relies on (3), (7). Since the definition
(10), (11) involves the highly non-linear operation $\oplus_q$ (7), problems arose upon a more careful examination of (7), not the least 
of which were questions about the domain of the employed functions/distributions and even more importantly for applications, 
about the invertibility of the $q$-Fourier transform.\\   

%%%%%%%%%%%%%%%%%%%%%%%%%%%%%%%%%%%%%%%%%%%%%%%%%%%%%%%%%%%%%%

\section{The $\mathbf{\tau_q}$-Fourier transform: motivation and definition} 

To avoid these issues some of which have been adequately resolved,   
we propose a different definition which is inspired and follows, up to a point, the spirit of equivariant  group actions on manifolds.
The advantage of following our proposed definition is that not only the $\tau_q$-Fourier transform is trivially invertible, 
but also it is unique under the assumption of exchange of point-wise products and convolutions, most surprisingly 
without the requirement of either linearity or continuity.  \\

The first step in this proposal is to abandon altogether the generalized binary operations (3), (7). They may be 
well-motivated and reasonable to define as well as have the nice properties (8), (9), but they have a main drawback 
in our opinion: the $q$-multiplication (7) is not distributive with respect to the $q$-addition (3). A comment may 
be in order about this: nature does not have to obey the dictates of existing mathematical structures. Actually many 
times the opposite is true, as modelling of natural phenomena has motivated the construction of deep and far-reaching 
mathematical structures. On the other hand, this has not happened very frequently over the years, so it may be 
prudent to adopt a more conservative approach and see whether any existing mathematical structures can be used to describe the phenomena 
under study and see how far they can lead us, before deciding to ignore them altogether, if needed.  \\

Given the above viewpoint, which is highly subjective, we would like to use the structure of a ring, field 
(in an algebraic sense) or module and vector space, if and whenever possible. This seems reasonable as physicists have considerable 
intuition and skills related to such structures, skills that have been honed over the many decades of their application to physical problems.
In our case, the most natural way to blend variants of (3) and (7) in an algebraic structure, is to use them to define a field.  
To attain this goal, we need to define operations that are distributive with respect to each other. This was done in \cite{PCPB}
and \cite{NK1}. To be more specific, in \cite{PCPB, NK1} the authors adopt the ``usual" definition for the $q$-sum  (3)
\begin{equation} 
      x\oplus_q y \ = \ x+y + (1-q) xy
\end{equation}
and the define $q$-product as 
\begin{equation}
      x \otimes_q y \ = \ \frac{1}{1-q} \left\{ (2-q)^\frac{\log [1+(1-q)x] \cdot \log [1+(1-q)y]}{[\log (2-q)]^2} - 1 \right\}
\end{equation}
To be fair, the  $q$-product (13) appeared only in \cite{PCPB}, with \cite{NK1} providing a definition based on an infinite 
series which is conjecturally equivalent to that of (13).  It is straightforward to check that (13) is distributive with respect to (12), namely  
\begin{equation}
    x \otimes_q (y \oplus_q z) \ = \ (x \otimes_q y) \oplus_q (x \otimes_q z)  
\end{equation}
Due to (14), a relabelled version of $\mathbb{R}$ endowed with (12), (13) becomes a field, which we called \ $\mathbb{R}_q$ \
in \cite{NK1}.  We found a field isomorphism \  $\tau_q$ \ between \ $\mathbb{R}$ \ and \  $\mathbb{R}_q$ \  which is explicitly given by 
\begin{equation}
     \tau_q(x) \ = \ \frac{1}{1-q} \{ (2-q)^x - 1\}
\end{equation}
Given this isomorphism, the generalized multiplication (13) which initially appears to be quite mysterious, looks practically trivial
\begin{equation}
    \tau_q (x \cdot y) \ = \ \tau_q \left( \tau_q^{-1}(x_q) \cdot \tau_q^{-1}(y_q) \right)
\end{equation}
or, even more pictorially, the following diagram commutes 
\begin{equation}
        \begin{CD}
         \mathbb{R} \times \mathbb{R}                            @ >  +, \ \cdot >>                                  \mathbb{R} \\
                    @V \tau_q \times \tau_q VV                                                                              @VV \tau_q V\\
         \mathbb{R}_q\times\mathbb{R}_q        @ >> \oplus_q, \ \otimes_q >         \mathbb{R}_q        
        \end{CD} 
\end{equation}  
where \ $x_q \equiv \tau_q(x)$. \ Then, all the algebraic relations of the Tsallis and the Boltzmann/Gibbs/ Shannon (BGS) 
entropies are mapped into each other via the isomorphism \ $\tau_q$. \\ 

\noindent Let the ordinary Fourier transform of a function \ $f:\mathbb{R} \rightarrow \mathbb{R}$ \ be indicated by \ $\mathbb{F}(f)$ \ namely 
\begin{equation}
   [\mathbb{F}(f)](k) \ = \ \int_\mathbb{R} e^{ikx} f(x) \ dx 
\end{equation}
One has to be careful with the functional space used in this definition: the actual requirement is for \ $f$ \ to be a distribution, 
something that will be more precisely dealt with at the end of Subsection 3.2, as such level of mathematical detail is only warranted 
 at that point.   We define the $\tau_q$-Fourier transform \ $[\mathfrak{F}_q(f_q)](\xi_q)$ \ 
 as the ordinary Fourier transform \ $\mathbb{F}(f)$ \ but defined in the field \ $\mathbb{R}_q$ \ 
 endowed with the  induced topology, limit, integral etc. from  \  $\mathbb{R}$ \ through \ $\tau_q$. \  
 In other words 
\begin{equation}
        [\mathfrak{F}_q (f_q)] (\xi_q) \ = \ \{ [\mathbb{F}(\tau_q (f))] (\tau_q (\xi)) \}_q 
\end{equation}
where the last subscript $q$ in the right-hand-side of (19) indicates that all calculations take place on and with the structures of 
\ $\mathbb{R}_q$. \ More pictorially, \ $\mathfrak{F}_q$ \ is defined in such a way that the following diagram commutes  
\begin{equation}
     \begin{CD}
          \mathbb{R}                            @ >  \tau_q >>                                  \mathbb{R}_q \\
                    @V \mathbb{F} VV                                                                              @VV \mathfrak{F}_q V\\
         \mathbb{R}                             @ > \tau_q ^\ast >>         \mathbb{R}_q        
        \end{CD} 
\end{equation}
In (20), \  $\tau_q ^\ast$ \ is the induced map from \ $\tau_q$ \ to the topological dual space of \ $\mathbb{R}$ \ with respect to the Fourier pairing,
the dual being also \ $\mathbb{R}$. \ Notice that diagram (20) is heuristic; its role is to convey the idea of the \ $\tau_q$-Fourier transform versus the 
usual Fourier transform \ $\mathbb{F}$, \ rather than being precise. \ 
In a more careful treatment, as in Subsection 3.2 below, let \ $\mathsf{S}_q$ \ and \ $\mathsf{S}^\prime _q$ \ 
indicate the spaces of Schwartz functions and of tempered distributions over \ $\mathbb{R}_q$ \ respectively. Then \ $\mathfrak{F}_q$ \
is defined so that the following diagram commutes
\begin{equation}
     \begin{CD}
          \mathsf{S}                            @ >  \tau_q >>                                  \mathsf{S}_q \\
                    @V \mathbb{F} VV                                                                              @VV \mathfrak{F}_q V\\
         \mathsf{S^\prime}                             @ > \tau_q ^\ast >>         \mathsf{S}^\prime _q        
        \end{CD} 
\end{equation}  
Expressed in terms of \ $\mathbb{R}$, \ the right-hand-side of (19) would be a rather messy and  
uninformative expression, so we refrain from explicitly showing the result.  \\

%%%%%%%%%%%%%%%%%%%%%%%%%%%%%%%%%%%%%%%%%%%%%%%%%%%%%%%%%%%%%%%%

\section{Covariance and uniqueness of the $\mathbf{\tau}_q$-Fourier transform}

\subsection{Covariance and the $\mathbf{\tau}_q$-Fourier transform} 

  What the definition (19) aims to do is to ascertain that the $\tau_q$-Fourier transform is equivariant under the action of \ $\tau_q$. \  
One reason for problems of (10) can be traced to the coexistence of the ordinary and the $q$-addition and $q$-multiplication
in the same expression. The latter (3), (7) are  non-linear expressions of elements of \ $\mathbb{R}$, \ and this non-linearity 
is the source of  many problems plaguing (10). The proposed definition (19) of the \ $\tau_q$-Fourier transform avoids this 
problem: because \ $\tau_q$ \ is a monotone map, the invertibility of the modified $\tau_q$-Fourier transform is trivially guaranteed. \\

Definitions like (19) are quite common in Mathematics and are also encountered  when one models physical systems by 
using group actions on manifolds such as homogeneous spaces, quotients of Lie groups etc. These are concretely encountered 
for configuration or phase spaces, for gauge field theoretical or general relativistic constructions etc 
possessing a high degree of symmetry which make explicit computations tractable.    
An important difference of all these treatments with the current one, is that as in the case 
of modern Algebraic Geometry (see, for instance, \cite{Harts}) following pioneering ideas and proposals  of  A. Grothendieck, 
one has to allow for changes of the underlying field in all constructions, thus needing definitions which are ``covariant" in a categorical sense. 
Indeed, there is suspicion that categorical constructions can be at the core of the formulation, and  may play a 
considerable  role in establishing and illuminating properties of the BGS entropy pertinent to physical systems \cite{Gr1, Gr2}. \\
   
The above is alongside the spirit  of the principle of general covariance which is the foundation of all gravitational theories today:
the underlying parametrization of a model of a physical system should be irrelevant for physically meaningful  results. 
Physical quantities must be observer-independent, even if their values depend on the observer: they must be covariant.  
The implementation of this principle plus the locality of the interactions, expressed through fields, necessitates the expression of 
physical quantities via tensorial constructions.   The underlying equations should be coordinate 
invariant, to express exactly this independence from specific non-physical parametrizations. This idea is a dominant theme and 
pervades especially gauge theories, General Relativity and Gravitation, String/M-theories etc. \\

Following this, a sketch of a rough analogy with General Relativity  and its terminology may be helpful in illustrating aspects of our viewpoint. 
Consider two persons (``observers"), the first  analyzing a system described by BGS entropy 
and the second analyzing a system described by the Tsallis entropy. These two systems are  physically quite different when seen by the same observer. 
Assume that the first person has at his disposal the algebraic structure of \ $\mathbb{R}$ \ and the second has at his disposal 
only \ $\mathbb{R}_q$. \  If we assume that the technique of choice for the analysis of these systems is the  Fourier transform, 
it would have to have a covariant definition to apply to both \ $\mathbb{R}$ \ and \ $\mathbb{R}_q$. \ The simplest way to do this, 
is to use the definition (19), in our opinion. If we ask each observer who lives in his own ``physical universe" 
(and they are not aware of the other observer) what they see, we would get the same answer:
each one of them sees a system that can be described by the BGS entropy but with respect to their respective algebraic fields 
\ $\mathbb{R}$ \  and \ $\mathbb{R}_q$. \  Then indeed the Fourier transform that they use would have the same form for each one 
observer.  The underlying Physics cannot be the same, of course,  as the BGS and the Tsallis entropies describe different physical phenomena. 
The discrepancy will become evident if the two persons compare non-algebraic aspects stemming from their distinct ground 
fields \ $\mathbb{R}$ \ and \  $\mathbb{R}_q$. \  Such a comparison will bring forth  the possible non-invertibility, lack of bijection, 
``singularities"  etc or other properties that the map \ $\tau_q$ \ would possess. 
Even though algebraically \ $\mathbb{R}$ \ and \ $\mathbb{R}_q$ \ are equivalent, the corresponding metric and measure 
(volume, or more general) structures mapped to each other via \  $\tau_q$ \ or maps induced by it, will not be equivalent. 
Therefore all physical differences between the two systems, when the two observers compare their 
analyses of their corresponding systems, should be ascribed  to non-trivial metric and measure theoretical 
properties induced by maps associated to \ $\tau_q$. \ This viewpoint has been at the core of our previous work on 
Tsallis entropy such as  \cite{NK2, NK3, NK4, NK5, NK6, NK7, NK8, NK9}. \\

In a way, what we do in the present work is to push the flexibility of defining the Fourier transform, from a plausible and educated or even ingenious, 
but largely arbitrary choice such as (10) or (11), to the flexibility of defining the appropriate isomorphism \ $\tau_q$. \  This approach is at least as old 
Abel's and Galois' ideas, where problems arising in algebraic structures such as polynomial rings are translated into problems of their 
automorphism groups.  Naturally our proposed modified $q$-Fourier transform (18) is not a unique choice: someone may decide 
to use the already existing $q$-Fourier or even define their own generalized Fourier transform tailored to their own needs and goals. 
However,  our proposal has an obvious advantage: if one adopts the underlying viewpoint about the
deformation from \  $\mathbb{R}$ \  to \  $\mathbb{R}_q$ \  through \ $\tau_q$ \ as encoding all the physical differences between the two systems, 
then once someone stays within \  $\mathbb{R}_q$ \ their choice of the Fourier transform is unique, under the assumption of the validity  
of two reasonable, and quite familiar to physicists, properties. \\       
   
%%%%%%%%%%%%%%%%%%%%%%%%%%%%%%   
   
\subsection{Uniqueness of the  $\mathbf{\tau_q}$-Fourier transform} 
 
The problem of what are the essential properties that are sufficient to uniquely characterize the ``ordinary" Fourier transform has a 
long history in a variety of contexts. Here we use the recent results of   
\cite{AAAM1, AAAM2, Jam, AAAFM} \ being the most pertinent for our purposes. 
These results followed the resolution of the same question for Legendre transforms \cite{AAKM, AAM}, a result whose 
physical implications for systems described by the Tsallis entropy we investigated in \cite{NK10}. \\
    
Even though the present work does not pretend to have  even a remote semblance of rigour up to this point,  we will attempt to be a bit more 
precise on the issue of uniqueness of the Fourier transform. This need is motivated by criticisms and viewpoints that gave rise to  
\cite{PR3, PR4, PR5} which call for a more precise treatment of some issues pertinent to some non-BGS (mainly the R\'{e}nyi and Tsallis) entropies. 
Moreover, someone may wish to revisit the commutative diagram (21) after reading this subsection.
On the rest of this subsection, we follow \cite{AAAM1, AAAM2, Jam} and especially  \cite{AAAFM}. 
For the required Fourier analysis background, one might wish to consult  \cite{SW, Stein}.   \\    

Consider the multi-index \ $\alpha = (\alpha_1, \ldots \alpha_n)$ \ where \ $\alpha_i \in \mathbb{Z}_+ \cup \{ 0 \}, \ i=1,\ldots, n$, indicate 
\  $|\alpha | = \alpha_1 + \ldots + \alpha_n$ \ and a \ $C^\infty$ \ function \ $f: \mathbb{R}^n \rightarrow \mathbb{C}$. \ Then $f$ is called a
Schwartz function, if for any \ $k\in \mathbb{Z}_+$, \  it satisfies    
\begin{equation}
     \sup_{x\in\mathbb{R}^n} \left| \frac{\partial^{|\alpha |} f(x)}{\partial x_1^{\alpha_1} \cdots \partial x_n^{\alpha_n}}  \left( 1 + |x|^k \right)   \right| \ < \infty
\end{equation}
The space of all such Schwartz functions \ $f$ \ will be indicated by \ $\mathsf{S}(n)$, \ or even more simply by \ $\mathsf{S}$. \  For \ $\mathsf{S}$, \ 
the  Fourier transform \ $\mathbb{F}: \mathsf{S} \rightarrow \mathsf{S}$ \ defined by 
\begin{equation}
      [\mathbb{F}(f)](\xi) \ = \ \int_{\mathbb{R}^n} f(x) e^{-2\pi i \langle x, \xi \rangle} \ dx 
\end{equation}
is a linear isomorphism, where \ $\langle\cdot, \cdot\rangle$ \ stands for the Euclidean inner product on \ $\mathbb{R}^n$. \  
The appearance of \ $2\pi$ \ in (23) when compared to (18) is just a normalization factor and is matter of convention.    
There are two  operations on \ $\mathsf{S}$: \ the (pointwise) product \ $\cdot$ \ and the convolution \ $\ast$ \ of functions. 
The topological dual space of  \ $\mathsf{S}$ \ is indicated by \ $\mathsf{S}^\prime$, \ is endowed with the weak topology 
and is called the space of tempered distributions. 
There is a canonical pairing \ $\mathsf{S} \rightarrow \mathsf{S}^\prime$, \ for \ $\phi\in\mathsf{S}, \  f\in\mathsf{S}^\prime$ 
\begin{equation}
    \langle \phi, f \rangle \ = \ \int_{\mathbb{R}^n} \phi(x) f(x) \ dx
\end{equation}
which is injective with a dense image. Under this pairing, there is the canonical inclusion \ $\mathsf{S} \rightarrow \mathsf{S}^\prime$. \
Then the Fourier transform \ $\mathbb{F}$ \ (23) induces an  isomorphism of \ $\mathsf{S}$ \ onto its image on \ $\mathsf{S}^\prime$ \ by
\begin{equation}
      F(\phi \cdot f) \ = \ F(\phi) \ast F(f)
\end{equation}
and 
\begin{equation}
     F (\phi \ast f) \ = \  F(\phi) \cdot F(f)
\end{equation}
These are the familiar properties of exchange of the multiplication of functions with convolution of their  Fourier transforms and vice-versa.
The key results of interest to us are: Assume that we have a bijection \ $\mathcal{F}: \mathsf{S} \rightarrow \mathsf{S}$ \ which admits  
a bijective extension \ $\mathcal{F}^\prime : \mathcal{S}^\prime \rightarrow \mathcal{S}^\prime $ \ such that \ 
$\mathcal{F}^\prime (\phi \cdot f) \ = \ \mathcal{F}(\phi) \ast \mathcal{F}^\prime (f)$. \ Then, there exists a diffeomorphism \ 
$h:\mathbb{R}^n \rightarrow \mathbb{R}^n$ \ such that for every \ $\phi\in\mathsf{S}$ \ either \ $\mathcal{F}(\phi) = \mathbb{F} (\phi  \circ h)$ \ or  \      
$\mathcal{F}(\phi) = {\mathbb{F}(\overline{\phi  \circ h})}$. \ A useful corollary of this result which characterizes the Fourier transform uniquely and even 
more explicitly is the following: Assume that \ $\mathcal{F},  \mathcal{F}^\prime$ \ and $\phi, f$ \ are defined as before, and are such that 
\ $\mathcal{F}^\prime (\phi \ast f)     \ = \ \mathcal{F}(\phi) \cdot \mathcal{F}^\prime (f)$ \ and \
 $\mathcal{F}^\prime (\phi \cdot f)   \ = \ \mathcal{F}(\phi) \ast \mathcal{F}^\prime (f)$. \ Then there is an \ $A\in GL_n(\mathbb{R})$, the linear group of 
 \ $n \times n$ \  matrices with real entries, \ with 
 \  $|\det (A)| = 1$ \ such that either \ $\mathcal{F} = \mathbb{F}(\phi \circ A)$ \ or \ $\mathcal{F} = \mathbb{F}(\overline{\phi \circ A})$ \ for all \ $\phi \in \mathsf{S}$. 
In other words, a transform exchanging pointwise multiplication with convolution is uniquely essentially the Fourier transform. 
A surprising observation is that in the above statements one assumes neither linearity nor continuity of \ $\mathcal{F}$. \  
These very well-known properties appear as a direct outcome of the assumptions alongside the definition of the Fourier transform.
From this particular viewpoint, nonlinear expressions such as the $q$-Fourier transform (10), (11) are un-natural and certainly not necessary.
It may also be worth noticing that the above statements can  be generalized to functions on smooth manifolds. When the above uniqueness result 
is combined with the monotonicity of \ $\tau_q$, \ then one sees that the proposed $\tau_q$-Fourier transform (19) is unique and invertible in a 
very natural way. Moreover, all the familiar properties of the classical Fourier transform (18) \cite{SW, Stein} carry over automatically to 
\ $\mathfrak{F}_q$ \ (19). \\  

%%%%%%%%%%%%%%%%%%%%%%%%%%%%%%%%%%%%%%%%%%%%%%%%%%%%%%%%%%%%%%%%%%%%%

\section{Conclusion and outlook}

We presented the motivation and the definition (19) of the $\tau_q$-Fourier transform. It is proposed as an attempt to sidestep some of the 
issues that have arisen in the definition and properties of the $q$-Fourier transform (10), (11) and its variants. The definition that we propose has the advantage 
of being unique, under  the exchange of convolutions with products, and trivially invertible. It is also linear for functions defined on its ground field \ $\mathbb{R}_q$ \ 
even if it appears highly non-linear and arcane from the viewpoint of \ $\mathbb{R}$. \ 
The definition (19) does not appear to have any obvious inconsistencies that would render it ill-defined. 
On the other  hand, its litmus test is how useful/effective such a construction  may prove to be in concrete calculations. 
This may be a worthy subject of investigation of some future work. \\

An advantage that definition (19) may have is its flexibility exactly due to its ``covariant" form. 
There are numerous entropic functionals other than the well-known  BGS and R\'{e}nyi ones, that have been proposed before and after the Tsallis entropy. 
The Tsallis/$q$- entropy is the best known and most developed non-BGS and non-R\'{e}nyi entropy in Physics. However, we  believe that it is unlikely that it 
will be the only one functional capable of  describing long spatial and temporal correlations, non-ergodic phase space behavior, systems out of equilibrium, 
systems with long-range interactions etc as many people claim.
We believe that these phenomena are too diverse for one non-trivial functional, such as the Tsallis/$q$- one,  to be able to capture all the possible statistics 
that is needed to describe the evolution and features of all such systems.\\

 Given the need for a multitude of entropies to describe different systems, 
one could repeat the same steps in determining the generalized sum and product as for Tsallis entropy for each of the other functionals. 
Then one could find the corresponding isomorphism like $\tau_q$, and then could easily formulate a corresponding generalized Fourier transform, 
like \ $\mathfrak{F}_q$ \ tailored to the properties of the entropy pertinent to the system under study. This was recently accomplished, for instance, 
for the \  $\kappa-$entropy in \cite{Scar1, Scar2}. In  \cite{Scar2} the exact analogue of (19) was proposed for the $\kappa$-entropy induced Fourier \
transform.  The constructions of \cite{Scar1, Scar2} rely on the substantial number of known properties induced by the $\kappa$-entropy \cite{Kaniad}.\\

Conversely, the arguments of the present work demonstrate that the $\kappa$-Fourier transform introduced \cite{Scar1, Scar2} is unique for the
$\kappa$-entropy induced non-additive statistical mechanics.  The mutually independent  but parallel developments
of the $\tau_q-$ and of the $\kappa-$ Fourier transforms may be seen as reinforcing the view that 
the path that we followed in constructing (19) as well as that of \cite{Scar2} may be effective and worthy of further exploration.\\

It is also evident through the present work, as well as that of \cite{Scar1, Scar2}, that if one defines differently the different generalized operations
then they will get different definitions, still unique within their underlying algebraic framework, of the generalized Fourier transform.  As was pointed out 
in  \cite{Scar3}, this is always possible when the theory is derived starting from a generalized trace-form entropic functional such as the $q$- or the
$\kappa$-entropies. The reason lies in the underlying ``covariance" of such definitions under the generalized operations, 
 which may be most clearly brought forth using the ``language" of category theory.\\ 

A contribution of the present work, as well as that of \cite{Scar1, Scar2}, may be the  realization that  definitions and properties 
formulated and proved in terms of category theory 
may be fruitful in defining and establishing properties related to the $\kappa$- , the $q$- or, more generally, to non-BGS entropies. 
This may provide an alternative framework, which is not necessarily mutually exclusive with the treatment of entropies 
via formal groups \cite{Temp1, Temp2}. 
The  viewpoints about the entropic functionals expressed via formal groups and categories may be useful
in establishing better commonalities and differences of such functionals at the formal level.\\ 

%%%%%%%%%%%%%%%%%%%%%%%%%%%%%%%%%%%%%%%%%%%%%%%%%%%%%%%%%%%%%%%%%%%%%%%%%%%%%%%%

%\vspace{10mm}

\noindent{\bf Acknowledgement:} \  We are grateful to the referees for suggesting improvements in the exposition and for making us aware of reference 
[43], and its meaning and significance.  We wish to thank Professor A. M. Scarfone for drawing our attention to his related work on $\kappa$-entropy and to references [40], [41].
We are  most grateful to Professor A. Bountis for his continuous encouragement and support in a multitude of ways.
We  are  grateful to Professors P. Benetatos,  I. Haranas  and  E.C. Vagenas for all their encouragement while this work was being carried out.\\

%%%%%%%%%%%%%%%%%%%%%%%%%%%%%%%%%%%%%%%%%%%%%%%%%%%%%%%%%%%%%%%%%%%%%%%%%%%%%%%%

%\newpage

%%%%%%%%%%%%%%%%%%%%%%%%%%%%%%%%%%%%%%%%%%%%%%%%%%%%%%%%%%%%%%%%%%%%

\end{document}